\documentclass[draftclsnofoot,onecolumn]{IEEEtran}
\usepackage{indentfirst,amsmath}
\usepackage{color,rotating,pdflscape,verbatim}
\usepackage{cite}
\usepackage{url}
\usepackage{mathtools}
\usepackage{amsfonts,amsmath,amssymb,amsthm}
\usepackage{latexsym,amscd}
\usepackage{amsbsy,extarrows}
\usepackage{enumitem}
\usepackage{graphicx,multirow,bm}
\usepackage{setspace}

\newtheorem{Theorem}{Theorem}

\newtheorem{Definition}{Definition}
\newtheorem{Example}{Example}

\newtheorem{Remark}{Remark}

\newtheorem{Proposition}{Proposition}

\newcommand{\CC}{\ensuremath{\mathbb{C}}}
\newcommand{\R}{\ensuremath{\mathbb{R}}}

\DeclareMathOperator{\Var}{Var}
\DeclareMathOperator{\Cov}{Cov}

\DeclareMathOperator{\Tr}{Tr}

\title{Analog Secure Distributed Matrix Multiplication over Complex Numbers}

\author{
  \IEEEauthorblockN{Okko Makkonen,~\IEEEmembership{Student Member,~IEEE} and Camilla Hollanti,~\IEEEmembership{Member,~IEEE}}

	\IEEEauthorblockA{Department of Mathematics and Systems Analysis\\ Aalto University, Finland\\ \texttt{\{okko.makkonen, camilla.hollanti\}@aalto.fi}
}}
\IEEEoverridecommandlockouts

\begin{document}

\maketitle
    
\begin{abstract}
This work considers the problem of distributing matrix multiplication over the real or complex numbers to helper servers, such that the information leakage to these servers is close to being information-theoretically secure. These servers are assumed to be honest-but-curious, \emph{i.e.}, they work according to the protocol, but try to deduce information about the data. The problem of secure distributed matrix multiplication (SDMM) has been considered in the context of matrix multiplication over finite fields, which is not always  feasible in real world applications. We present two schemes, which allow for variable degree of security based on the use case and allow for colluding and straggling servers. We analyze the security and the numerical accuracy of the schemes and observe a trade-off between accuracy and security.  
\end{abstract}
    
\section{Introduction}\label{sec:intro}

Secure distributed matrix multiplication has been researched as a way to harness the power of distributed computation for large scale matrix multiplication. Different code constructions and fundamental limits have been studied in \cite{chang2018capacity, chang2019upload, jia2019capacity, yang2019secure, d2020gasp, d2021degree, d2020notes, jia2021cross, mital2020secure, kim2019privateCL, yu2020entangled, yu2019lagrange, kakar2019capacity}, which are discussed in more detail in Section \ref{sec:related_work}. When distributing computation to multiple workers, the computation time will be limited by the slowest one. This limitation can be quite severe, since the system might have stragglers that take much longer time than the others. To mitigate this straggler-effect, coded computation is utilized. Coded computation relies on coding theoretic tools that allow recovering the final result from only a partial set of answers. This way the user only needs to wait for the fastest responses before getting the final result. Furthermore, distributing sensitive data to unknown helpers might not be feasible in some circumstances, \textit{e.g.}, sensitive medical information or identifiable personal data. To protect the confidentiality of the information, tools from information theory and coding theory are used. This allows the scheme to be secure against any $X$ colluding servers.

Existing SDMM schemes compute the product of matrices over finite fields, since coding theory is most often done over finite fields. However, computing a matrix product over a finite field is not as practical as computing a product over the real or complex numbers. It is possible to discretize matrix multiplication over the real numbers to a suitable prime field, which loses some precision. Furthermore, the operations over a finite field are slow when compared to operations over reals when approximating with floating point numbers. The approximation done by floating point numbers is much better than the approximation done by converting to finite fields, since floating point numbers do not have to trade magnitude with precision.

\subsection{Related Work}\label{sec:related_work}

Distributed matrix multiplication with different code constructions have been studied in \cite{lee2017speeding, joshi2017efficient, wang2015using, yu2017polynomial, dutta2018unified, yu2020straggler, dutta2020optimal, dutta2019optimal}. Secure distributed matrix multiplication was first studied in \cite{chang2018capacity}, where the outer product partitioning was used (\emph{cf.} Section \ref{sec:scheme_construction}). This was improved in \cite{d2020gasp, d2021degree, d2020notes} by using GASP codes. Different codes have also been introduced in \cite{kakar2019capacity, aliasgari2020private, jia2019capacity, mital2020secure, kim2019privateCL, yu2020entangled}. Different modes of SDMM, such as private or batch SDMM, have been studied in \cite{chang2019upload, jia2021cross, yu2019lagrange, yu2020entangled, chen2021gcsa, zhu2021secure}. The capacity of SDMM has been studied in \cite{chang2018capacity, jia2019capacity, kakar2019capacity, yang2019secure} in some special cases, but the general problem remains open.

The problem of secure coded computation over the analog domain was considered in \cite{tjell2021privacy, soleymani2020privacy, soleymani2021analog}. Secret sharing over real numbers is considered in \cite{tjell2021privacy} using Lagrange coded secret sharing. In \cite{soleymani2020privacy, soleymani2021analog} coded distributed polynomial evaluation over matrices is considered over the complex numbers.

\subsection{Motivation and Contribution}\label{sec:motivation}

There is only one problem one initially faces when moving from a finite field to the real numbers. All existing SDMM schemes sample matrices from a uniform distribution over the entire field. As this is not possible over an infinite field, some other distribution has to be used. Other than this, most existing schemes can be computed using real numbers and the computation will be correct. The important things to consider are the security of the scheme and the numerical stability.

In this paper we present two SDMM schemes that work over the complex numbers and show that they can achieve information leakage that is arbitrarily small by a suitable choice of the parameters. These schemes have been previously considered in the context of SDMM over finite fields in \cite{aliasgari2020private, d2020gasp, d2021degree}, but this paper provides a more practical way of using them in the real world. A similar idea of using roots of unity as evaluation points was recently considered in \cite{soleymani2020privacy, soleymani2021analog} in the context of coded distributed polynomial evaluation. This work combines earlier work on SDMM over finite fields and the work of \cite{soleymani2020privacy, soleymani2021analog} to a novel and practical coded computation scheme. 

\section{Preliminaries}\label{sec:preliminaries}

\subsection{General SDMM Model}\label{sec:general_model}

The user is interested in computing the product of matrices $A$ and $B$ of size $t \times s$ and $s \times r$, respectively. We consider these matrices to be over $\R$ or $\CC$. The user does this computation using the help of $N$ honest-but-curious servers. Any set of at most $X$ servers is allowed to collude, \emph{i.e.}, share all their data to try to learn information about the matrices $A$ and $B$.

The general computational model for SDMM schemes is the following.
\begin{itemize}
    \item Encoding phase: The user partitions the matrices to smaller submatrices and draws random matrices of the same size. These smaller matrices are encoded using some code and the encoded pieces $\widetilde{A}_i$, $\widetilde{B}_i$ are sent to server $i$. This part can be seen as a secret sharing phase, where the secret matrices are shared among the servers.
    \item Computation phase: Each server computes the product of their encoded pieces and returns the result $\widetilde{A}_i \widetilde{B}_i$ to the user. The responses can be seen as shares of another secret sharing scheme, where the secret contains the wanted result.
    \item Decoding phase: Using the results of the fastest servers the user is able to decode the final result $AB$ by computing a certain linear combination (or linear combinations) of the responses.
\end{itemize}

The security of an SDMM scheme is usually defined as perfect information theoretic security, since that is available in finite fields. Over an infinite field it is not possible to obtain perfect information theoretic security, but using the following definition it is possible to limit the amount of leaked information with the parameter $\delta > 0$. The parameter $\delta$ can be chosen relative to the entropy of the matrices $A$ and $B$.

\begin{Definition}\label{def:delta_security}
An SDMM scheme is $(X, \delta)$-secure for $\delta > 0$ if
\begin{equation*}
    I(\mathbf{\mathbf{A}, \mathbf{B}; \widetilde{A}}_\mathcal{X}, \mathbf{\widetilde{B}}_\mathcal{X}) \leq \delta
\end{equation*}
for all $\mathcal{X} \subset [N]$, $|\mathcal{X}| = X$. Here $\mathbf{\widetilde{A}}_\mathcal{X}$ denotes $\{ \mathbf{\widetilde{A}}_i \}_{i \in \mathcal{X}}$, and the bold variables denote the random variables and nonbold denote the specific values of those random variables.
\end{Definition}

There are many interesting metrics when comparing SDMM schemes, such as upload cost, download cost and recovery threshold. In this paper we are not interested in comparing these parameters as we are presenting a new type of scheme that works over a different field.

\subsection{Numerical Stability}\label{sec:numerical_stability}

When working with finite field arithmetic, there are no considerations of numerical inaccuracies, since finite field elements can be represented exactly on a computer. However, this is not the case for real or complex numbers, which are usually represented using some kind of floating point representation. The limitations of the finite precision of the floating point system introduce errors in the computation, which might grow quickly if the computational task is not well-conditioned.

The decoding process of an SDMM scheme involves inverting a submatrix of the generator matrix. In its most general form this can be represented as solving the system of linear equations defined by $Ax = y$ where $A \in \CC^{n \times n}$ and $y \in \CC^n$ are known. The solution $x$ will be erroneous if there are errors in $y$. The error can be measured using the absolute error $\lVert \delta x \rVert$ or the relative error $\frac{\lVert \delta x \rVert}{\lVert x \rVert}$, where $\delta x = x - \hat{x}$ is the error, $\hat{x}$ is the computed value and $x$ is the exact value. The condition number $\kappa(A)$ of the matrix describes how much larger the relative error in $x$ will be compared to the relative error in $y$. The condition number is bounded from below by 1 and a small condition number is desirable for a stable computation.

Measuring the error of some matrix operation can be done by computing some matrix norm of the difference between the real and computed results. Such matrix norms include the Frobenius norm. For more information on numerical stability of algorithms we refer to \cite{higham2002accuracy}.

\subsection{Random Variables and Maximal Entropy}\label{sec:maximal_entropy}

For random variables over finite fields the entropy is maximized by a uniform random variable. Therefore, it is advantageous to use uniform distributions to hide information in a coded computation setting. Some other distribution needs to be used when working in an infinite field such as the real or complex numbers.

In this paper, we work with continuous random variables that have a probability density function. The covariance matrix of a real random vector is denoted by $\Sigma_X$ and its $(i, j)$ entry is $\Cov(\mathbf{X}_i, \mathbf{X}_j) = \mathbb{E}[(\mathbf{X}_i - \mathbb{E}[\mathbf{X}_i])(\mathbf{X}_j - \mathbb{E}[\mathbf{X}_j])^*]$. A complex random variable $\mathbf{Z}$ is \textit{circular} if $\mathbf{Z} \sim e^{i\theta} \mathbf{Z}$ for any deterministic $\theta \in \R$. In particular we can consider a circular complex Gaussian random vector, which is characterized by its covariance matrix. We can formulate the following proposition about maximal entropy distributions over the real and complex numbers.

\begin{Proposition}\label{prop:maximal_entropy}
\begin{enumerate}[label=(\roman*)]
    \item \cite[Theorem 8.6.5]{cover1999elements} Let $\mathbf{X}$ be a continuous real random vector of length $n$ with a nonsingular covariance matrix $\Sigma$. Then
    \begin{equation*}
        h(\mathbf{X}) \leq \tfrac{1}{2}\log((2\pi e)^n \det(\Sigma))
    \end{equation*}
    with equality if and only if $\mathbf{X}$ is a Gaussian random vector.
    \item \cite[Theorem 2]{neeser1993proper} Let $\mathbf{Z}$ be a continuous complex random vector of length $n$ with a nonsingular covariance $\Sigma$. Then
    \begin{equation*}
        h(\mathbf{Z}) \leq \log((\pi e)^n \det(\Sigma))
    \end{equation*}
    with equality if and only if $\mathbf{Z}$ is a circular complex Gaussian random vector.
\end{enumerate}
\end{Proposition}

\subsection{Secret Sharing over the Real Numbers}\label{sec:secret_sharing_reals}

Secret sharing is a way of distributing a secret to $n$ parties such that some subset of those parties can use their shares to reconstruct the secret. A $(k, n)$-threshold secret sharing scheme allows for any $k$ parties to recover the secret, which is usually considered to be a finite field value. Secret sharing over the real numbers was recently considered in \cite{tjell2021privacy}. The following example shows how such a scheme works and how the security is determined.

\begin{Example}\label{ex:secret_sharing_reals}
Consider a $(2, 3)$-threshold secret sharing scheme, where any 2 of the 3 parties can recover the secret. Let $a \in \R$ be the secret. The shares are constructed with
\begin{equation*}
    \widetilde{A}_i = a + r\alpha_i,
\end{equation*}
for $i=1, 2, 3$, where $r \in \R$ is drawn at random from some distribution independently of $a$. Without loss of generality, assume that parties $1$ and $2$ want to recover the secret from their shares $\widetilde{A}_1$ and $\widetilde{A}_2$. Consider the linear combination
\begin{equation*}
    \alpha_2\widetilde{A}_1 - \alpha_1\widetilde{A}_2 = (\alpha_2 - \alpha_1)a + (\alpha_2 \alpha_1 - \alpha_1 \alpha_2)r = (\alpha_2 - \alpha_1)a.
\end{equation*}
Then the secret $a$ can be recovered, provided that the points $\alpha_1, \alpha_2, \alpha_3$ are distinct.

Let us now analyze the amount of information leaked from one share. To do this we consider the values $a$ and $r$ to come from some probability distributions as the values of the random variables $\mathbf{a}$ and $\mathbf{r}$, respectively. Similarly, the share $\widetilde{A}_i$ is the value of the random variable $\mathbf{\widetilde{A}}_i$. Using the definition of mutual information we get that
\begin{align*}
    I(\mathbf{\widetilde{A}}_i; \mathbf{a}) &= h(\mathbf{\widetilde{A}}_i) - h(\mathbf{\widetilde{A}}_i \mid \mathbf{a}) 
    = h(\mathbf{a} + \mathbf{r} \alpha_i) - h(\mathbf{a} + \mathbf{r} \alpha_i \mid \mathbf{a}) \\
    &= h(\mathbf{a} + \mathbf{r} \alpha_i) - h(\mathbf{r}\alpha_i).
\end{align*}
The last step follows from the independence of $\mathbf{a}$ and $\mathbf{r}$. Assuming that the distribution of $\mathbf{a}$ is continuous, we can use Proposition \ref{prop:maximal_entropy}(i). The variance of the sum is then $\Var(\mathbf{a} + \mathbf{r}\alpha_i) = \sigma_s^2 + \alpha_i^2\sigma_r^2$. Therefore,
\begin{equation*}
    h(\mathbf{a} + \mathbf{r} \alpha_i) \leq \tfrac{1}{2}\log(2\pi e (\sigma_a^2 + \alpha_i^2\sigma_r^2)).
\end{equation*}
We may choose that $\mathbf{r}$ is distributed according to a Gaussian distribution with mean 0 and variance $\sigma_r^2$. Therefore,
\begin{equation*}
    h(\mathbf{r} \alpha_i) = \tfrac{1}{2}\log(2 \pi e \alpha_i^2 \sigma_r^2).
\end{equation*}
Hence, the amount of leaked information from share $\mathbf{\widetilde{A}}_i$ is
\begin{equation*}
    I(\mathbf{\widetilde{A}}_i; \mathbf{a}) \leq \frac{1}{2} \log(1 + \frac{\sigma_a^2}{\alpha_i^2 \sigma_r^2}).
\end{equation*}
By growing $\sigma_r^2$ we can make the amount of leaked information as small as we wish. It is also clear that we cannot have the situation that $\alpha_i = 0$, since then the share $\widetilde{A}_i$ would leak the secret directly.
\end{Example}

It is worth noticing that the amount of leaked information depends on the value of $\alpha_i^2$, which means that some shares leak more information than others. The computation in the example is just an upper bound, but an explicit number can be computed if the distribution of $\mathbf{a}$ is known. If $\mathbf{a}$ distributed according to a Gaussian distribution, then the upper bound is reached with equality by Proposition \ref{prop:maximal_entropy}(i).

The encoding and recovery phases of SDMM schemes can be seen as secret sharing schemes where the secrets and random parts are matrices. This is in principle similar to the example above, but the dependencies between the different components need to be taken into account.

The problem with this type of secret sharing scheme is that the amount of information leaked depends on the evaluation points. In the above example we use the basis $1, x, x^2, \dots$, which introduces the problem that some of the random values are multiplied with larger constants, which makes them contribute more to the entropy. In \cite{tjell2021privacy} the basis is chosen to be Lagrange basis polynomials. The structure of the existing SDMM schemes relies heavily on the basis that is used, so for SDMM it is not practical to change the basis polynomials to Lagrange basis polynomials. In this work we fix the problem by extending the scheme to the complex numbers and choosing the evaluation points from the unit circle.

\newpage
\section{Scheme Constructions}\label{sec:scheme_construction}

In this section we present two explicit analog SDMM schemes based on previous schemes over finite fields. 

\subsection{Secure MatDot Code over Complex Numbers}\label{sec:analog_matdot}

The MatDot code over finite fields was first introduced in \cite{dutta2019optimal} without the security constraint and later in \cite{aliasgari2020private} with the security constraint. The following construction extends the scheme to the complex numbers.

The input matrices are $A \in \CC^{t \times s}$ and $B \in \CC^{s \times r}$. These matrices come from some continuous probability distribution over the complex numbers. The matrices are then split to $p$ pieces, assuming that $p \mid s$. Hence, the partition is the \emph{inner product partition} (IPP)
\begin{equation*}
    A = \begin{psmallmatrix}
    A_1 & \dots & A_p
    \end{psmallmatrix}, \quad B = \begin{psmallmatrix}
    B_1 \\ \vdots \\ B_p
    \end{psmallmatrix}.
\end{equation*}
The partitions are of the size $A_j \in \CC^{t \times \frac{s}{p}}$ and $B_{j'} \in \CC^{\frac{s}{p} \times r}$. The product $AB$ is then the dot product of the partitions, \textit{i.e.},
\begin{equation*}
    AB = \sum_{j=1}^p A_j B_j.
\end{equation*}
Matrices $R_1, \dots, R_X \in \CC^{t \times \frac{s}{p}}$ and $S_1, \dots, S_X \in \CC^{\frac{s}{p} \times r}$ are drawn at random such that each component is distributed according to a circular complex Gaussian distribution with zero mean and variance $\sigma^2$. The partitions and the random matrices are encoded using the polynomials
\begin{align*}
    f(x) &= \sum_{j=1}^p A_j x^{j-1} + \sum_{k=1}^X R_k x^{p + k - 1} \\
    g(x) &= \sum_{j'=1}^p B_{j'} x^{p-j'} + \sum_{k'=1}^X S_{k'} x^{p + k' - 1}. 
\end{align*}
Let $N$ be the number of servers to use. Then choose the evaluation points as $N$th roots of unity, \textit{i.e.}, $\alpha_i = \zeta_N^i$, where $\zeta_N$ is the primitive $N$th root of unity. Then the evaluations $\widetilde{A}_i = f(\alpha_i)$ and $\widetilde{B}_i = g(\alpha_i)$ are sent to server $i$. Each server multiplies their encoded matrices and returns the product $\widetilde{A}_i \widetilde{B}_i$. These results are evaluations of the polynomial $h(x) = f(x)g(x)$. Using the definition of $f(x)$ and $g(x)$ we get that
\begin{align*}
    & h(x) = f(x)g(x) \\
    &= \sum_{j=1}^p \sum_{j'=1}^p A_j B_{j'}x^{p - 1 + j - j'} + \sum_{j=1}^p \sum_{k'=1}^X A_j S_{k'} x^{p + j + k' - 2} \\
    &+ \sum_{k=1}^X \sum_{j'=1}^p R_k B_{j'}x^{2p - 1 + k - j'} + \sum_{k=1}^X \sum_{k'=1}^X R_k S_{k'} x^{2p + k + k' - 2}.
\end{align*}
The coefficient of the term $x^{p-1}$ is $\sum_{j=1}^p A_j B_j$, which equals the product $AB$. The degree of $h(x)$ is $2p + 2X - 2$, so the user needs $2p + 2X - 1$ evaluations to be able to interpolate the result and choose the coefficient of the term $x^{p-1}$.

In practice, the interpolation would be done by inverting the Vandermonde matrix associated with the evaluation points $\alpha_i$ for $i \in \mathcal{I}$, where $\mathcal{I} \subset [N]$ is the set of $K = 2p + 2X - 1$ fastest servers. The inverse Vandermonde matrix would then be multiplied with the vector of responses such that the $(p-1)$th element of the resulting vector is the product $AB$. For numerical stability it is important that the condition number of that Vandermonde matrix is small.

\subsection{GASP Code over Complex Numbers}

The following scheme follows the construction of the $\mathrm{GASP}_\text{big}$ code presented in \cite{d2020gasp}. We use this special case of the more general GASP code presented in \cite{d2021degree}, since the $\mathrm{GASP}_\text{big}$ code is a subcode of a Generalized Reed--Solomon code with a small dimension. Therefore, it is possible to use a Vandermonde matrix as the generator matrix, which makes it simple to choose the evaluation points of the scheme.

Again, the input matrices are $A \in \CC^{t \times s}$ and $B \in \CC^{s \times r}$, which come from some continuous probability distribution over the complex numbers. The matrices are then split to $m$ and $n$ pieces, respectively, assuming that $m \mid t$ and $n \mid r$. Hence, the partition is the \emph{outer product partition} (OPP)
\begin{equation*}
    A = \begin{psmallmatrix}
    A_1 \\ \vdots \\ A_m
    \end{psmallmatrix}, \quad B = \begin{psmallmatrix}
    B_1 & \dots & B_n
    \end{psmallmatrix}.
\end{equation*}
The partitions are of the size $A_j \in \CC^{\frac{t}{m} \times s}$ and $B_{j'} \in \CC^{s \times \frac{r}{n}}$. The product $AB$ is then the dot product of the partitions, \textit{i.e.},
\begin{equation*}
    AB = \begin{psmallmatrix}
    A_1B_1 & \dots & A_1B_n \\
    \vdots & & \vdots \\
    A_mB_1 & \dots & A_mB_n
    \end{psmallmatrix}.
\end{equation*}
Matrices $R_1, \dots, R_X \in \CC^{\frac{t}{m} \times s}$ and $S_1, \dots, S_X \in \CC^{s \times \frac{r}{n}}$ are drawn at random such that each component is distributed according to a circular symmetric complex Gaussian distribution with zero mean and variance $\sigma^2$. The partitions and the random matrices are encoded using the polynomials
\begin{align*}
    f(x) &= \sum_{j=1}^p A_j x^{j-1} + \sum_{k=1}^X R_k x^{mn + k - 1} \\
    g(x) &= \sum_{j'=1}^p B_{j'} x^{m(j' - 1)} + \sum_{k'=1}^X S_{k'} x^{mn + k' - 1}. 
\end{align*}
The evaluation points $\alpha_1, \dots, \alpha_N$ are again chosen as $N$th roots of unity and each server gets the evaluation of $f(x)$ and $g(x)$ at the evaluation point. The responses, after the servers have multiplied their encoded pieces, are evaluations of $h(x) = f(x)g(x)$. Using the definition of $f(x)$ and $g(x)$ from above we get
\begin{align*}
    & h(x) = f(x)g(x) \\
    &= \sum_{j=1}^m \sum_{j'=1}^n A_j B_{j'} x^{m(j' -1) + j - 1} + \sum_{j=1}^m \sum_{k'=1}^X A_j S_{k'} x^{mn + k' + j - 2} \\
    &+ \sum_{k=1}^X \sum_{j'=1}^n R_k B_{j'} x^{mn + m(j' - 1) + k - 1} + \sum_{k=1}^X \sum_{k'=1}^X R_k S_{k'} x^{2mn + k + k' - 2}.
\end{align*}
It can now be seen that the coefficient of $x^{m(j' - 1) + j - 1}$ is exactly $A_jB_{j'}$ for $(j, j') \in [m] \times [n]$. Therefore, the product $AB$ can be recovered from the first $mn$ coefficients of $h(x)$. The degree of $h(x)$ is $2mn + 2X - 2$, so the user needs $2mn + 2X - 1$ evaluations to recover the result.

The interpolation is done again by multiplying by the inverse of a corresponding Vandermonde matrix, which means that it is important to have a small condition number.

\subsection{Analysis of Security}\label{sec:security}

We shall only prove the security of the analog MatDot scheme as the proof for the analog GASP scheme is analogous.

We can write the encoding process of the analog MatDot scheme by using the generator matrices $F$ and $G$, respectively, which contain the relevant coefficients used in the encoding. Then we get
\begin{align*}
    \widetilde{A} &= (\widetilde{A}_1, \dots, \widetilde{A}_N) = (A_1, \dots, A_p, R_1, \dots, R_X)F, \\
    \widetilde{B} &= (\widetilde{B}_1, \dots, \widetilde{B}_N) = (B_1, \dots, B_p, S_1, \dots, S_X)G,
\end{align*}
for some generator matrices $F$ and $G$. These can be written as
\begin{align*}
    \widetilde{A} &= \underbrace{(A_1, \dots, A_p)F^{\leq p}}_{=A'} + \underbrace{(R_1, \dots, R_X)F^{> p}}_{=R'}, \\
    \widetilde{B} &= \underbrace{(B_1, \dots, B_p)G^{\leq p}}_{=B'} + \underbrace{(S_1, \dots, S_X)G^{> p}}_{=S'},
\end{align*}
where $F^{\leq p}$ and $F^{> p}$ denote the first $p$ rows and the last $X$ rows of $F$, respectively. We wish to now analyze how much information is leaked to any set of $X$ colluding servers. We wish to limit the amount of leaked information to some $\delta > 0$, by setting the variance of the random matrices suitably.

Let $\mathcal{X} \subset [N]$ be the indices of the colluding servers with $|\mathcal{X}| = X$. Then
\begin{align*}
    I(\mathbf{A}, \mathbf{B}; \mathbf{\widetilde{A}}_\mathcal{X}, \mathbf{\widetilde{B}}_\mathcal{X}) &= I(\mathbf{A}, \mathbf{B}; \mathbf{\widetilde{A}}_\mathcal{X}) + I(\mathbf{A}, \mathbf{B}; \mathbf{\widetilde{B}}_\mathcal{X} \mid \mathbf{\widetilde{A}}_\mathcal{X}) \\
    &= h(\mathbf{\widetilde{A}}_\mathcal{X}) - h(\mathbf{\widetilde{A}}_\mathcal{X} \mid \mathbf{A}, \mathbf{B}) \\
    &\quad + h(\mathbf{\widetilde{B}}_\mathcal{X} \mid \mathbf{\widetilde{A}}_\mathcal{X}) - h(\mathbf{\widetilde{B}}_\mathcal{X} \mid \mathbf{\widetilde{A}}_\mathcal{X}, \mathbf{A}, \mathbf{B}) \\
    &\leq h(\mathbf{\widetilde{A}}_\mathcal{X}) - h(\mathbf{\widetilde{A}}_\mathcal{X} \mid \mathbf{A}) \\
    &\quad + h(\mathbf{\widetilde{B}}_\mathcal{X}) - h(\mathbf{\widetilde{B}}_\mathcal{X} \mid \mathbf{B}) \\
    &= I(\mathbf{A}; \mathbf{\widetilde{A}}_\mathcal{X}) + I(\mathbf{B}; \mathbf{\widetilde{B}}_\mathcal{X}).
\end{align*}
The inequality follows from (1) $\mathbf{\widetilde{A}}_\mathcal{X}$ is conditionally independent of $\mathbf{B}$ given $\mathbf{A}$, (2) conditioning lowers the entropy, and (3) $\mathbf{\widetilde{B}}_\mathcal{X}$ is conditionally independent of $\mathbf{\widetilde{A}}_\mathcal{X}$ and $\mathbf{A}$ given $\mathbf{B}$. Without loss of generality, let us analyze the first term, since the proof is the same for the second term. We start by writing
\begin{equation*}
    I(\mathbf{A}; \mathbf{\widetilde{A}}_\mathcal{X}) = h(\mathbf{\widetilde{A}}_\mathcal{X}) - h(\mathbf{\widetilde{A}}_\mathcal{X} \mid \mathbf{A}).
\end{equation*}
Let us start by analyzing the first term. Let $\mathbf{\widetilde{A}}$ denote some element of the random matrix $\mathbf{\widetilde{A}}$ and similarly $\mathbf{\widetilde{A}}_\mathcal{X}$ be the random vector containing some element of the random matrix for all $i \in \mathcal{X}$. From the union bound on entropy we get that
\begin{align*}
    h(\mathbf{\widetilde{A}}_\mathcal{X}) &\leq \sum h(\mathbf{\widetilde{A}}_\mathcal{X}) 
    \leq \sum \log((\pi e)^X \det(\Sigma_{\widetilde{A}})) \\
    &= \frac{ts}{p} \log((\pi e)^X \det(\Sigma_{\widetilde{A}})),
\end{align*}
where the sums are over all the elements in the encoded matrix. We used Proposition \ref{prop:maximal_entropy}(ii), since $\mathbf{\widetilde{A}}_\mathcal{X}$ is a continuous complex random variable. For simplicity we assume that all elements are identically distributed, which allows us to consider some element of the $t \times \frac{s}{p}$ matrix. For the second term we get
\begin{align*}
    h(\mathbf{\widetilde{A}_\mathcal{X}} \mid \mathbf{A}) = h(\{ \mathbf{A}'_i + \mathbf{R}'_i \}_{i \in \mathcal{X}} \mid \mathbf{A}).
\end{align*}
Now, $\mathbf{A}'_i$ is completely determined by $\mathbf{A}$, so the conditional entropy is determined directly by the entropy of the random matrices. Then, we can again separate the components of the random matrices, since all the components are independent. Hence, we get
\begin{equation*}
    h(\mathbf{\widetilde{A}_\mathcal{X}} \mid \mathbf{A}) = \frac{ts}{p} h(\mathbf{r}'_\mathcal{X}) = \frac{ts}{p} \log((\pi e)^X \det(\Sigma_{r'})),
\end{equation*}
by Proposition \ref{prop:maximal_entropy}(ii). Therefore,
\begin{equation*}
    I(\mathbf{A}; \mathbf{\widetilde{A}}_\mathcal{X}) \leq \frac{ts}{p}\log\left(\frac{\det(\Sigma_{\widetilde{A}})}{\det(\Sigma_{r'})}\right).
\end{equation*}
We now need to set an upper bound for the quotient of the determinants. Let us analyze the covariance matrices. By independence of $\mathbf{a}'_\mathcal{X}$ and $\mathbf{r}'_\mathcal{X}$, we can write $\Sigma_{\widetilde{A}} = \Sigma_{a'} + \Sigma_{r'}$. As $\Sigma_{a'}$ is positive semi-definite, we can write $\Sigma_{a'} = VV^*$ for some $V \in \CC^{X \times X}$. By definition
\begin{align*}
    \Sigma_{r'} = \Var(rF^{> p}_{\mathcal{X}}) &= F^{> p}_\mathcal{X} \Var(r) (F^{> p}_\mathcal{X})^* \\
    &= \sigma_r^2 F^{> p}_\mathcal{X} (F^{> p}_\mathcal{X})^*,
\end{align*}
since $\Var(r) = \sigma_r^2 I_X$ as the random matrices are chosen independently with the same variance. Define $\Gamma_a = F^{> p}_\mathcal{X} (F^{> p}_\mathcal{X})^*$, which is invertible, since $F^{> p}_\mathcal{X}$ is a Vandermonde matrix multiplied with an invertible diagonal matrix. We now use the matrix determinant lemma to compute $\det(\Sigma_{\widetilde{A}})$. Then we get
\begin{equation*}
    \det(\Sigma_{\widetilde{A}}) = \det(\Sigma_{r'} + \Sigma_{a'}) = \det(\Sigma_{r'}) \det(I_m + \frac{1}{\sigma_r^2} V^*\Gamma_a^{-1} V).
\end{equation*}
Now, $I_X + \frac{1}{\sigma_r^2} V^*\Gamma_a^{-1} V$ is positive semi-definite. Thus, we can use Hadamard's inequality to get
\begin{equation*}
    \det(\Sigma_{\widetilde{A}}) \leq \det(\Sigma_{r'}) \prod_{i=1}^p \left( 1 + \frac{1}{\sigma_r^2} (V^*\Gamma_a^{-1} V)_{ii} \right).
\end{equation*}
Therefore, we get
\begin{align*}
    I(\mathbf{\widetilde{A}}_{\mathcal{X}}; \mathbf{A}) &\leq \frac{ts}{p} \log\left( \prod_{i=1}^p \left( 1 + \frac{1}{\sigma_r^2} (V^*\Gamma_a^{-1} V)_{ii} \right) \right) \\
    &= \frac{ts}{p} \sum_{i=1}^p \log\left( 1 + \frac{1}{\sigma_r^2} (V^*\Gamma_a^{-1} V)_{ii} \right) \\
    &\leq \frac{ts}{p} \frac{1}{\ln 2} \frac{1}{\sigma_r^2} \Tr(V^*\Gamma_a^{-1} V) \\
    &= \frac{ts}{p \ln 2} \frac{1}{\sigma_r^2} \Tr(\Gamma_a^{-1} \Sigma_{a'}).
\end{align*}

We can now make the leaked information as small as we wish by growing $\sigma_r^2$. We summarize the result in the following theorem.

\begin{Theorem}\label{thm:security_variance_bound}
The analog MatDot scheme with parameters $p$ and $X$ achieves $(X, \delta)$-security by choosing the variance such that
\begin{equation*}
    \sigma^2 \geq \max_{\stackrel{\mathcal{X} \subset [N]}{|\mathcal{X}| = X}}\frac{s}{\delta p \ln 2}\left( t \Tr(\Gamma_a^{-1} \Sigma_{a'}) + r \Tr(\Gamma_b^{-1} \Sigma_{b'}) \right).
\end{equation*}
\end{Theorem}

\begin{Remark}
To compute $\Sigma_{a'}$ or $\Sigma_{b'}$ we need to know the covariance matrix of the elements of $A$ or $B$. This is simple in the case that the elements are i.i.d., but complicated when this is not the case. It is an interesting future research question to construct an upper bound even if the covariance matrix is not fully known.
\end{Remark}

Here $\Gamma_a$ and $\Sigma_{a'}$ depend on the colluding set $\mathcal{X}$, which means that we need to consider the maximal value of the trace to get an upper bound that is simple to compute, without needing to check all possible colluding sets. We conjecture that the trace is maximized when the evaluation points are chosen as consecutive roots of unity around the unit circle.

\subsection{Analysis of Numerical Accuracy}\label{sec:accuracy}

\begin{figure}[t]
    \centering
    \includegraphics[width=0.6\textwidth]{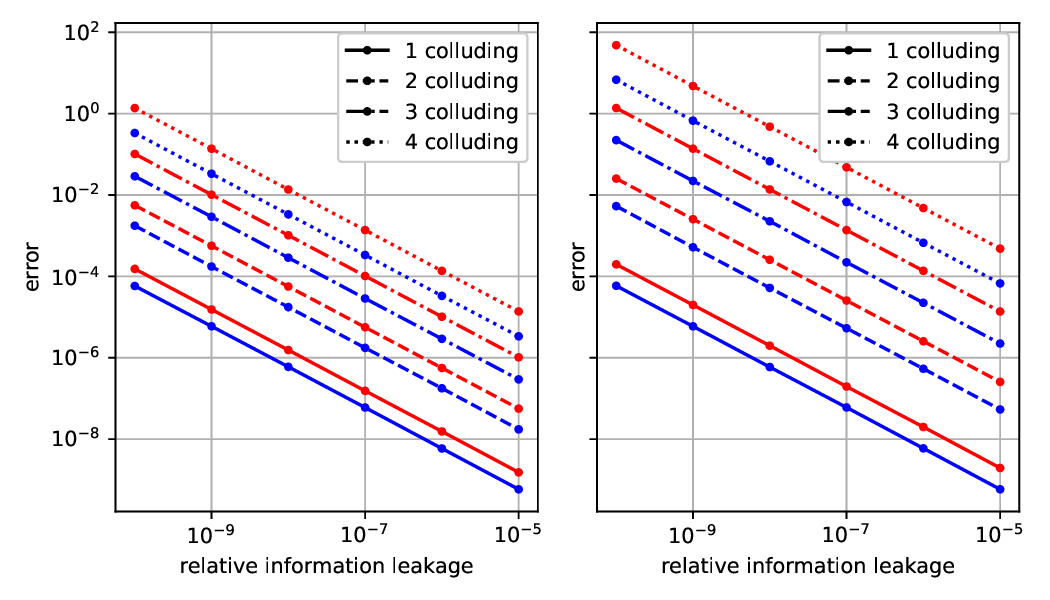}
    \caption{Comparison of the effect of colluding servers on the relationship between the leaked information and the numerical error. The blue lines correspond to MatDot with partitioning parameter $p = 4$ (left) and $p = 9$ (right). The red lines correspond to the analog GASP scheme with partitioning parameters $m = n = 2$ (left) and $m = n = 3$ (right).}
    \label{fig:collusion}
\end{figure}

\begin{figure}[t]
    \centering
    \includegraphics[width=0.6\textwidth]{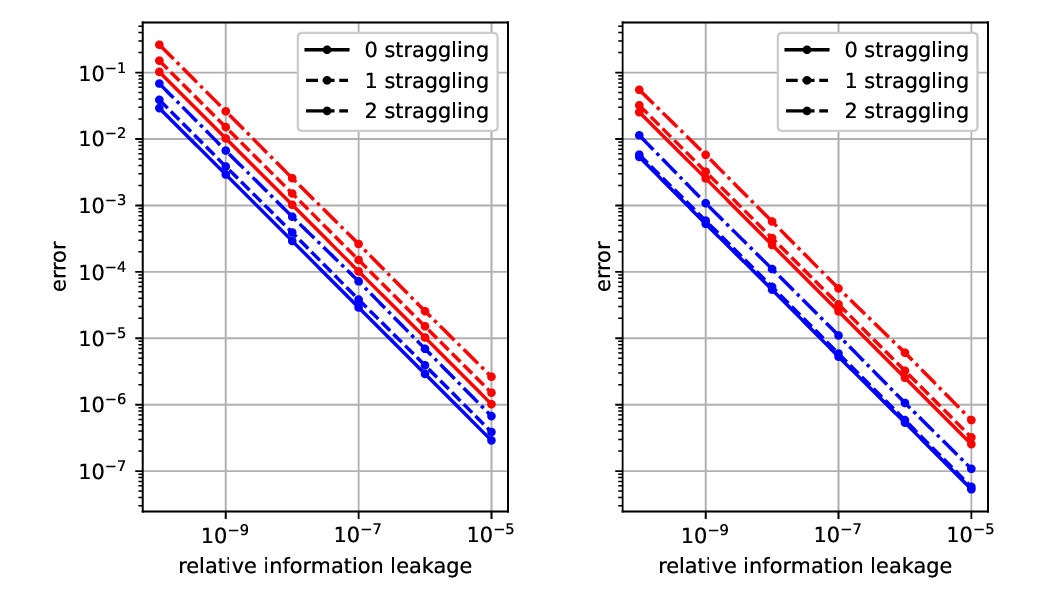}
    \caption{Comparison of the effect of straggling servers on the relationship between the leaked information and the numerical error. The blue lines correspond to MatDot with partitioning parameter $p = 4$ (left) and $p = 9$ (right). The red lines correspond to the analog GASP scheme with partitioning parameters $m = n = 2$ (left) and $m = n = 3$ (right). The number of colluding servers is $X = 3$ (left) and $X = 2$ (right).}
    \label{fig:straggling}
\end{figure}

To make the computation secure, we need to make the variance $\sigma^2$ sufficiently large. By doing this, we add large numbers to our original input data, which means that the elements in the encoded matrices are on average large compared to the unencoded matrices. As the numerical error of matrix multiplication is proportional to the sizes of the elements, the size of the error compared to the original data grows as the variance grows. This introduces a trade-off between security and numerical accuracy. 

We analyze the additional error introduced by the SDMM process by running the algorithm 1000 times and recording the mean error as a function of the specified security level. For each round new $36 \times 36$ input matrices are chosen, such that each element is independently chosen from the standard Gaussian distribution. The variance is chosen to be as low as possible using Theorem \ref{thm:security_variance_bound}. The resulting graphs are plotted in Figures \ref{fig:collusion} and \ref{fig:straggling}. The error is measured as the Frobenius norm of the difference of the product as computed by the SDMM scheme compared to the product as compared to the computation done without SDMM. The experiments were performed using IEEE double precision floating point numbers. The leaked information is measured as the relative information loss compared to the entropy of the input matrices $\mathbf{A}$ and $\mathbf{B}$.\footnote{The source code for the generation of the plots can be found on \\ \url{https://github.com/okkomakkonen/sdmm-over-complex}.}

From each of the figures it is clear that there is an inverse relationship between the security and the numerical accuracy. This is explained by the fact that for large enough values of variance, the input data, which is small in comparison, is irrelevant, which means that the standard deviation $\sigma$ scales the magnitude of the encoded matrices. As the numerical error of matrix multiplication is directly proportional to the product of the magnitudes of the inputs, the error is directly proportional to the variance $\sigma^2$. This doesn't apply for low values of the variance, where the actual input data also has an affect, which can also be confirmed numerically. 

From Figure \ref{fig:collusion} it is clear that growing the number of colluding servers also increases the numerical error. The increase seems to be linear in $X$, which is expected, since for large values of the variance the number of random matrices determines the size of the elements in the encoded matrices. 

In the case that there are no stragglers, the evaluation points are chosen as the roots of unity, which means that the corresponding Vandermonde matrix used in the decoding process is unitary. Hence, the condition number is $1$. In \cite{gautschi1990stable} it was shown that for real evaluation points the condition number grows exponentially with $N$, which means that using only real evaluation points would not be numerically stable. In the case of straggling servers the Vandermonde matrix will not be unitary, since it does not contain all of the $N$th roots of unity. In \cite{ramamoorthy2019numerically} it was shown that the condition number grows polynomially in the number of servers, when the number of stragglers is constant. This behaviour can be seen in Figure \ref{fig:straggling}, where the numerical error grows as the number of straggling servers grows.

From Figures \ref{fig:collusion} and \ref{fig:straggling} it is clear that the analog MatDot code achieves better numerical accuracy compared to the analog GASP code when $p = mn$ and the other parameters are the same. However, the analog GASP code achieves a better download cost by using the outer product partitioning.

\section*{Acknowledgements}

This work has been supported by the Academy of Finland, under Grants No. 318937 and 336005. The authors would like to thank Dr.\ Diego Villamizar Rubiano for useful discussions.

\bibliography{arxiv.bib}

\begin{thebibliography}{10}
\providecommand{\url}[1]{#1}
\csname url@samestyle\endcsname
\providecommand{\newblock}{\relax}
\providecommand{\bibinfo}[2]{#2}
\providecommand{\BIBentrySTDinterwordspacing}{\spaceskip=0pt\relax}
\providecommand{\BIBentryALTinterwordstretchfactor}{4}
\providecommand{\BIBentryALTinterwordspacing}{\spaceskip=\fontdimen2\font plus
\BIBentryALTinterwordstretchfactor\fontdimen3\font minus
  \fontdimen4\font\relax}
\providecommand{\BIBforeignlanguage}[2]{{%
\expandafter\ifx\csname l@#1\endcsname\relax
\typeout{** WARNING: IEEEtran.bst: No hyphenation pattern has been}%
\typeout{** loaded for the language `#1'. Using the pattern for}%
\typeout{** the default language instead.}%
\else
\language=\csname l@#1\endcsname
\fi
#2}}
\providecommand{\BIBdecl}{\relax}
\BIBdecl

\bibitem{chang2018capacity}
W.-T. Chang and R.~Tandon, ``On the capacity of secure distributed matrix
  multiplication,'' in \emph{2018 IEEE Global Communications Conference
  (GLOBECOM)}.\hskip 1em plus 0.5em minus 0.4em\relax IEEE, 2018, pp. 1--6.

\bibitem{chang2019upload}
------, ``On the upload versus download cost for secure and private matrix
  multiplication,'' in \emph{2019 IEEE Information Theory Workshop
  (ITW)}.\hskip 1em plus 0.5em minus 0.4em\relax IEEE, 2019, pp. 1--5.

\bibitem{jia2019capacity}
Z.~Jia and S.~A. Jafar, ``On the capacity of secure distributed matrix
  multiplication,'' \emph{arXiv preprint arXiv:1908.06957}, 2019.

\bibitem{yang2019secure}
H.~Yang and J.~Lee, ``Secure distributed computing with straggling servers
  using polynomial codes,'' \emph{IEEE Transactions on Information Forensics
  and Security}, vol.~14, no.~1, pp. 141--150, 2019.

\bibitem{d2020gasp}
R.~G. D'Oliveira, S.~El~Rouayheb, and D.~Karpuk, ``{GASP} codes for secure
  distributed matrix multiplication,'' \emph{IEEE Transactions on Information
  Theory}, vol.~66, no.~7, pp. 4038--4050, 2020.

\bibitem{d2021degree}
R.~G. D’Oliveira, S.~El~Rouayheb, D.~Heinlein, and D.~Karpuk, ``Degree tables
  for secure distributed matrix multiplication,'' \emph{IEEE Journal on
  Selected Areas in Information Theory}, vol.~2, no.~3, pp. 907--918, 2021.

\bibitem{d2020notes}
R.~G. D'Oliveira, S.~E. Rouayheb, D.~Heinlein, and D.~Karpuk, ``Notes on
  communication and computation in secure distributed matrix multiplication,''
  in \emph{2020 IEEE Conference on Communications and Network Security
  (CNS)}.\hskip 1em plus 0.5em minus 0.4em\relax IEEE, 2020, pp. 1--6.

\bibitem{jia2021cross}
Z.~Jia and S.~A. Jafar, ``Cross subspace alignment codes for coded distributed
  batch computation,'' \emph{IEEE Transactions on Information Theory}, vol.~67,
  no.~5, pp. 2821--2846, 2021.

\bibitem{mital2020secure}
N.~Mital, C.~Ling, and D.~Gunduz, ``Secure distributed matrix computation with
  discrete {Fourier} transform,'' \emph{arXiv preprint arXiv:2007.03972}, 2020.

\bibitem{kim2019privateCL}
M.~Kim and J.~Lee, ``Private secure coded computation,'' \emph{IEEE
  Communications Letters}, vol.~23, no.~11, pp. 1918--1921, 2019.

\bibitem{yu2020entangled}
Q.~Yu and A.~S. Avestimehr, ``Entangled polynomial codes for secure, private,
  and batch distributed matrix multiplication: Breaking the `cubic' barrier,''
  in \emph{2020 IEEE International Symposium on Information Theory
  (ISIT)}.\hskip 1em plus 0.5em minus 0.4em\relax IEEE, 2020, pp. 245--250.

\bibitem{yu2019lagrange}
Q.~Yu, S.~Li, N.~Raviv, S.~M.~M. Kalan, M.~Soltanolkotabi, and S.~A.
  Avestimehr, ``Lagrange coded computing: Optimal design for resiliency,
  security, and privacy,'' in \emph{The 22nd International Conference on
  Artificial Intelligence and Statistics}.\hskip 1em plus 0.5em minus
  0.4em\relax PMLR, 2019, pp. 1215--1225.

\bibitem{kakar2019capacity}
J.~Kakar, S.~Ebadifar, and A.~Sezgin, ``On the capacity and
  straggler-robustness of distributed secure matrix multiplication,''
  \emph{IEEE Access}, vol.~7, pp. 45\,783--45\,799, 2019.

\bibitem{lee2017speeding}
K.~Lee, M.~Lam, R.~Pedarsani, D.~Papailiopoulos, and K.~Ramchandran, ``Speeding
  up distributed machine learning using codes,'' \emph{IEEE Transactions on
  Information Theory}, vol.~64, no.~3, pp. 1514--1529, 2017.

\bibitem{joshi2017efficient}
G.~Joshi, E.~Soljanin, and G.~Wornell, ``Efficient redundancy techniques for
  latency reduction in cloud systems,'' \emph{ACM Transactions on Modeling and
  Performance Evaluation of Computing Systems (TOMPECS)}, vol.~2, no.~2, pp.
  1--30, 2017.

\bibitem{wang2015using}
D.~Wang, G.~Joshi, and G.~Wornell, ``Using straggler replication to reduce
  latency in large-scale parallel computing,'' \emph{ACM SIGMETRICS Performance
  Evaluation Review}, vol.~43, no.~3, pp. 7--11, 2015.

\bibitem{yu2017polynomial}
Q.~Yu, M.~Maddah-Ali, and S.~Avestimehr, ``Polynomial codes: an optimal design
  for high-dimensional coded matrix multiplication,'' in \emph{Advances in
  Neural Information Processing Systems}, 2017, pp. 4403--4413.

\bibitem{dutta2018unified}
S.~Dutta, Z.~Bai, H.~Jeong, T.~M. Low, and P.~Grover, ``A unified coded deep
  neural network training strategy based on generalized {PolyDot} codes,'' in
  \emph{2018 IEEE International Symposium on Information Theory (ISIT)}.\hskip
  1em plus 0.5em minus 0.4em\relax IEEE, 2018, pp. 1585--1589.

\bibitem{yu2020straggler}
Q.~Yu, M.~A. Maddah-Ali, and A.~S. Avestimehr, ``Straggler mitigation in
  distributed matrix multiplication: Fundamental limits and optimal coding,''
  \emph{IEEE Transactions on Information Theory}, vol.~66, no.~3, pp.
  1920--1933, 2020.

\bibitem{dutta2020optimal}
S.~Dutta, M.~Fahim, F.~Haddadpour, H.~Jeong, V.~Cadambe, and P.~Grover, ``On
  the optimal recovery threshold of coded matrix multiplication,'' \emph{IEEE
  Transactions on Information Theory}, vol.~66, no.~1, pp. 278--301, 2020.

\bibitem{dutta2019optimal}
------, ``On the optimal recovery threshold of coded matrix multiplication,''
  \emph{IEEE Transactions on Information Theory}, vol.~66, pp. 278--301, 2019.

\bibitem{aliasgari2020private}
M.~Aliasgari, O.~Simeone, and J.~Kliewer, ``Private and secure distributed
  matrix multiplication with flexible communication load,'' \emph{IEEE
  Transactions on Information Forensics and Security}, vol.~15, pp. 2722--2734,
  2020.

\bibitem{chen2021gcsa}
Z.~Chen, Z.~Jia, Z.~Wang, and S.~A. Jafar, ``{GCSA} codes with noise alignment
  for secure coded multi-party batch matrix multiplication,'' \emph{IEEE
  Journal on Selected Areas in Information Theory}, vol.~2, no.~1, pp.
  306--316, 2021.

\bibitem{zhu2021secure}
J.~Zhu and X.~Tang, ``Secure batch matrix multiplication from grouping
  {Lagrange} encoding,'' \emph{IEEE Communications Letters}, vol.~25, no.~4,
  pp. 1119--1123, 2021.

\bibitem{tjell2021privacy}
K.~Tjell and R.~Wisniewski, ``Privacy in distributed computations based on real
  number secret sharing,'' \emph{arXiv preprint arXiv:2107.00911}, 2021.

\bibitem{soleymani2020privacy}
M.~Soleymani, H.~Mahdavifar, and A.~S. Avestimehr, ``Privacy-preserving
  distributed learning in the analog domain,'' \emph{arXiv preprint
  arXiv:2007.08803}, 2020.

\bibitem{soleymani2021analog}
------, ``Analog {Lagrange} coded computing,'' \emph{IEEE Journal on Selected
  Areas in Information Theory}, vol.~2, no.~1, pp. 283--295, 2021.

\bibitem{higham2002accuracy}
N.~J. Higham, \emph{Accuracy and stability of numerical algorithms}.\hskip 1em
  plus 0.5em minus 0.4em\relax SIAM, 2002.

\bibitem{cover1999elements}
T.~M. Cover, \emph{Elements of information theory}.\hskip 1em plus 0.5em minus
  0.4em\relax John Wiley \& Sons, 1999.

\bibitem{neeser1993proper}
F.~D. Neeser and J.~L. Massey, ``Proper complex random processes with
  applications to information theory,'' \emph{IEEE transactions on information
  theory}, vol.~39, no.~4, pp. 1293--1302, 1993.

\bibitem{gautschi1990stable}
W.~Gautschi, ``How (un) stable are {Vandermonde} systems?'' in \emph{Asymptotic
  and computational analysis}.\hskip 1em plus 0.5em minus 0.4em\relax CRC
  Press, 1990, pp. 193--210.

\bibitem{ramamoorthy2019numerically}
A.~Ramamoorthy and L.~Tang, ``Numerically stable coded matrix computations via
  circulant and rotation matrix embeddings,'' \emph{arXiv preprint
  arXiv:1910.06515}, 2019.

\end{thebibliography}
\bibliographystyle{IEEEtran}

\end{document}